\documentclass[]{aastex631}

\usepackage{graphicx}
\usepackage{xspace}

\newcommand{\package}[1]{\texttt{#1}}
\newcommand{\sorcha}{\package{Sorcha}\xspace}
\newcommand{\python}{\package{Python}\xspace}

\usepackage{calc}
\newlength{\okinalen}
\newcommand{\oumuamua}{\hbox to.666\okinalen{\hss`\hss}Oumuamua}

\def \MIDAS {Michigan Institute for Data and AI in Society, University of Michigan, Ann Arbor, MI 48109, USA}
\def \Physics {Department of Physics, University of Michigan, Ann Arbor, MI 48109, USA}

\def \CfA {Center for Astrophysics | Harvard \& Smithsonian, 60 Garden Street, Cambridge, MA 02138, USA}

\begin{document}

\title{Sorcha: Optimized Solar System Ephemeris Generation}

\author[0000-0002-1139-4880]{Matthew~J.~Holman}
\affil{\CfA}
\correspondingauthor{Matthew~J.~Holman}
\email{mholman@cfa.harvard.edu}

\author[0000-0003-0743-9422]{Pedro H. Bernardinelli} 
\affiliation{DiRAC Institute and the Department of Astronomy, University of Washington, 3910 15th Ave NE, Seattle, WA 98195, USA}

\author[0000-0003-4365-1455]{Megan E. Schwamb}
\affiliation{Astrophysics Research Centre, School of Mathematics and Physics, Queen's University Belfast, Belfast BT7 1NN, UK}

\author[0000-0003-1996-9252]{Mario Juri\'c}
\affiliation{DiRAC Institute and the Department of Astronomy, University of Washington, 3910 15th Ave NE, Seattle, WA 98195, USA}

\author[0000-0001-6984-8411]{Drew Oldag}
\affiliation{DiRAC Institute and the Department of Astronomy, University of Washington, 3910 15th Ave NE, Seattle, WA 98195, USA}
\affiliation{LSST Interdisciplinary Network for Collaboration and Computing Frameworks, 933 N. Cherry Avenue, Tucson AZ 85721}

\author[0009-0003-3171-3118]{Maxine West}
\affiliation{DiRAC Institute and the Department of Astronomy, University of Washington, 3910 15th Ave NE, Seattle, WA 98195, USA}
\affiliation{LSST Interdisciplinary Network for Collaboration and Computing Frameworks, 933 N. Cherry Avenue, Tucson AZ 85721}

\author[0000-0003-4827-5049]{Kevin~J.~Napier}
\affiliation{\MIDAS}
\affiliation{\Physics}
\affiliation{\CfA}

\author[0000-0001-5930-2829]{Stephanie R. Merritt}
\affiliation{Astrophysics Research Centre, School of Mathematics and Physics, Queen's University Belfast, Belfast BT7 1NN, UK}

\author[0000-0002-8418-4809]{Grigori Fedorets}
\affiliation{Astrophysics Research Centre, School of Mathematics and Physics, Queen's University Belfast, Belfast BT7 1NN, UK}
\affiliation{Finnish Centre for Astronomy with ESO, University of Turku, FI-20014 Turku, Finland}
\affiliation{Department of Physics, University of Helsinki, P.O. Box 64, 00014
Helsinki, Finland}

\author[0000-0002-0672-5104]{Samuel Cornwall} 
\affiliation{Department of Aerospace Engineering, University of Illinois at Urbana-Champaign,
Urbana, IL 61801, USA}

\author[0009-0005-5452-0671]{Jacob A. Kurlander} 
\affiliation{DiRAC Institute and the Department of Astronomy, University of Washington, 3910 15th Ave NE, Seattle, WA 98195, USA}

\author[0000-0002-1398-6302]{Siegfried Eggl}
\affiliation{Department of Aerospace Engineering,
University of Illinois at Urbana-Champaign,
Urbana, IL 61801, USA}
\affiliation{Department of Astronomy,
University of Illinois at Urbana-Champaign,
Urbana, IL 61801, USA}
\affiliation{National Center for Supercomputing Applications,
University of Illinois at Urbana-Champaign,
Urbana, IL 61801, USA}

\author[0009-0009-2281-7031]{Jeremy Kubica}
\affiliation{McWilliams Center for Cosmology, Department of Physics, Carnegie Mellon University, Pittsburgh, PA 15213, USA}
\affiliation{LSST Interdisciplinary Network for Collaboration and Computing Frameworks, 933 N. Cherry Avenue, Tucson AZ 85721}

\author{Kathleen Kiker}
\affiliation{Asteroid Institute, 20 Sunnyside Ave., Suite 427, Mill Valley, CA 94941, USA}

\author[0000-0001-9505-1131]{Joseph Murtagh}
\affiliation{Astrophysics Research Centre, School of Mathematics and Physics, Queen's University Belfast, Belfast BT7 1NN, UK}

\author[0000-0003-4439-7014]{Shantanu P. Naidu}
\affiliation{Jet Propulsion Laboratory, California Institute of Technology, Pasadena, CA, USA}

\author[0000-0001-7335-1715]{Colin Orion Chandler}
\affiliation{DiRAC Institute and the Department of Astronomy, University of Washington, 3910 15th Ave NE, Seattle, WA 98195, USA}
\affiliation{LSST Interdisciplinary Network for Collaboration and Computing Frameworks, 933 N. Cherry Avenue, Tucson AZ 85721}

\begin{abstract}
\sorcha is a solar system survey simulator built for the Vera C. Rubin Observatory's Legacy Survey of Space and Time (LSST) and future large-scale wide-field surveys. Over the ten-year survey, the LSST is expected to collect roughly a billion observations of minor planets. The task of a solar system survey simulator is to take a set of input objects (described by orbits and physical properties) and determine what a real or hypothetical survey would have discovered. Existing survey simulators have a computational bottleneck in determining which input objects lie in each survey field, making them infeasible for LSST data scales. \sorcha can swiftly, efficiently, and accurately calculate the on-sky positions for sets of millions of input orbits and surveys with millions of visits, identifying which exposures these objects cross, in order for later stages of the software to make detailed estimates of the apparent magnitude and detectability of those input small bodies. In this paper, we provide the full details of the algorithm and software behind \sorcha's ephemeris generator. Like many of \sorcha's components, its ephemeris generator can be easily used for other surveys.
\end{abstract}

\section{Introduction}\label{sec:intro}

With operations scheduled to start in the second half of 2025, the Vera C. Rubin Observatory's Legacy Survey of Space and Time  \citep[LSST; ][]{lsst-sciencebook-ch5-2009,ivezic2019,bianco2022}  will map $\sim 18,000 \, \deg^2$ of the sky continuously for 10 years, with roughly 2 million total visits expected.
The survey will use the LSST Camera (LSSTCam), with a field of view of $9.6 \, \deg^2$, a typical expected depth of $m_r \approx 24.0$ per $30\, \mathrm{sec}$ visit, and an expected astrometric noise floor of 10 mas.
LSST's combination of area, cadence, depth, and astrometric precision will enable it to discover and subsequently track almost five to ten times more objects than are currently known across all Solar System populations, from near-Earth objects (NEOs) to main belt asteroids (MBAs) to objects in the distant Solar System \citep[][;Murtagh et al. (submitted); Kurlander et al (submitted)]{2009EM&P..105..101J, lsst-sciencebook-ch5-2009, 2010Icar..205..605S, 2015MNRAS.446.2059S, 2016AJ....152..103S, 2016AJ....151..172G, 2017AJ....154...13V, 2018Icar..303..181J,  ivezic2019, 2020Icar..33813517F, 2022PSJ.....3...71H}. 

A number of excellent survey simulators have been developed for previous solar system surveys~\citep{Petit.2011,Naidu.2017,Napier.2021,Bernardinelli.2022,Bernardinelli.2024}, but they are not ideally suited to the needs of the LSST; they either do not scale well to the size of LSST or they make approximations that limit their precision in ways that are incompatible with the community's main use cases in the LSST era \citep{2018arXiv180201783S}.  \sorcha is a survey simulator designed and built specifically to tackle the challenges of simulating which objects in a model solar system population the LSST would discover~(Merritt et al. 2025, submitted). 
\sorcha takes an input solar system population  (a set of synthetic objects with orbits, absolute magnitudes, colors, and phase curve parameters) and determines which of the input objects would be detectable in an input survey's pointings. Once there is a set of potential detections per survey pointing, \sorcha then replicates the various interlinked observational biases to estimate which of these potential detections can be successfully detected by the survey source detection algorithms and subsequently identified by the survey's moving object discovery algorithms.\footnote{Merritt et al 2025 submitted, provides an in-depth introduction to how the simulator is built and works. The source code is written in \texttt{Python}, and the repository and documentation are accessible online.  Repository: \url{https://github.com/dirac-institute/sorcha} and Documentation: \url{https://sorcha.readthedocs.io}. }
In this paper, we describe the ephemeris generation algorithm powering \sorcha, which quickly and efficiency identifies for each of the exposures which of the input orbits land within a circular approximation of the survey's camera's footprint and then calculates their precise on-sky locations at the mid-point of the observation.  

As Rubin is expected to discover objects across all Solar System populations, approximations such as using Keplerian two-body solutions, or even treating only the giant planets as perturbers (as is common for outer Solar System objects), are not sufficient, as such approximations provide enough precision for minor body identification and long term tracking given Rubin's expected astrometric performance.
Furthermore LSST's discoveries will be made with data spanning weeks, months, and years.
Thus, \sorcha needs to propagate a large sample of orbits through the full set of survey exposures in order to simulate discovery process, which is a significant computational challenge. 
We confine ourselves to finding an efficient solution to this problem: Given a sample of orbits of solar system objects, defined at a reference epoch, and a set of visits (exposure times and telescope pointings), find the sets of objects that appear within the field-of-view (FOV) of the camera at each of the visits. Although simulating Rubin's LSST is our primary motivation, our solution to this problem is widely applicable.

The remainder of this paper is organized as follows.  In section \ref{sec:simple}, we describe the process of generating an ephemeris for a single solar system body observed at a single time from a single observatory.  This serves to illustrate the essential steps in the calculation.  We describe how we make those individual steps more efficient.  We also compare our ephemeris results to those of other ephemeris generation tools, to ensure that they are correct.
In section \ref{sec:large-scale}, we describe our solution to enabling ephemeris generation at LSST scale.  
Next, in section~\ref{sec:validation}, we test our optimizations by comparing the results of our solution to those of a brute force application of the technique described in section~\ref{sec:simple}.  In section~\ref{sec:caveats}, we discuss a number of caveats related to our solution to large-scale ephemeris generation. Finally, we summarize our results and suggest possible future work in section~\ref{sec:summary}.

\section{Simple Ephemeris Generation} \label{sec:simple}

In this section,  we describe each of the steps in generating an ephemeris position for a single small solar system body, observed at a single time.  This illustrates the computational costs of each of the steps and motivates the design choices within \sorcha that we make to optimize this process for a significant population of objects calculated a large number of times.

For our purposes, an ephemeris position is a unit vector from the observatory in the direction of the solar system body at the time of the observation:
\begin{equation}
\mathbf{\hat{\rho}}(t) = \frac{\mathbf{r}(t-\Delta t) - \mathbf{r_{obs}}(t)}{|\mathbf{r}(t-\Delta t) - \mathbf{r_{obs}}(t)|},
\label{Eq:light-time}
\end{equation}
where $\mathbf{\hat{\rho}}(t)$ is the observed unit vector at time $t$, $\mathbf{r}(t-\Delta t)$ is the position of the object at time $t-\Delta t$, and $\mathbf{r_{obs}}(t)$ is the position of the observatory at time $t$.  Both the position of the object and the position of the observatory are measured from the same origin (the solar system barycenter).  The light time is $\Delta t = |\mathbf{r}(t-\Delta t) - \mathbf{r_{obs}}(t)|/c$, where $c$ is the speed of light. 

There are a few primary calculations needed to generate an ephemeris position with Equation~\ref{Eq:light-time}:
(1) determining the location of the observatory at the time of the observation; 
(2) converting the orbit to initial conditions suitable for numerical integration; 
(3) determining the location of the object via numerical integration;
and (4) adjusting that location to account for light time.
We describe and validate each of these steps and the overall process in the following subsections. 

\subsection{Observatory Location}

Determining the location of the observatory is critical to obtaining accurate ephemeris positions; if the position of the observatory is poorly known, the calculated unit vector toward the object will be as well~(Eq.~\ref{Eq:light-time}).
For a ground-based observatory, one calculates the position of the geocenter with respect to the Sun or barycenter, calculates the position of the observatory with respect to the geocenter, and then sums the two.

While the process of calculating the observatory position can be computationally expensive, an observatory's position only needs to be calculated once per survey exposure.  These calculations are done in advance within \sorcha. A table of the locations of ground-based observatories is maintained by the Minor Planet Center (MPC)\footnote{The MPC provides this information in the observatory codes data file available at \url{https://www.minorplanetcenter.net/data}}.  These are geocentric positions with respect to the Earth's body-fixed frame.  One must carry out a rotation, or a sequence of rotations, to get the position of the observatory with respect to the geocenter in an inertial frame.  Traditionally, this was done with formulas that expressed the time-dependent direction of the Earth's pole and its rate of rotation of the Earth~\citep{Urban.2014}.  However, the tidal influence of the Moon and Sun, as well as the time-dependent moments of inertia of the Earth, imply that the rotation state of the Earth can only be predicted approximately.  
Instead, models of the Earth's rotation state are fit to observations to precisely determine its past values and to provide accurate predictions for the Earth's rotation state.  The model fits and predictions are updated daily.  In addition to the positions of the planets, NASA's Navigation and Ancillary Information Facility provides SPICE\footnote{Spacecraft Planet Instrument C-matrix Events}  kernels that tabulate the rotation matrix from the body-fixed frame of the Earth (ITRF93) to the International Celestial Reference Frame (ICRF), as a function of time. Specifically, we use the SPICE function {\it pxform} to obtain the matrix for this rotation, for each observation at the exposure mid-times.  
We verify our calculation by comparing the results to those from JPL's Horizons service\footnote{\url{https://ssd.jpl.nasa.gov/horizons/app.html}} and find agreement at the level of $\sim10$~m.

\subsection{Initial Conditions and Coordinate Conversations} \label{sec:initial_conditions}

As input for numerical integration, we need equatorial barycentric state vectors $\{\mathbf{r}, \dot{\mathbf{r}}\}$. However, as an input, we wish to be able to handle sets of (ecliptic-aligned) Keplerian elements (semi-major axis $a$, eccentricity $e$, inclination $i$, longitude of ascending node $\Omega$, argument of pericenter $\omega$, and mean anomaly $\mathcal{M}$) or cometary orbital elements (pericenter distance $q$, eccentricity $e$, inclination $i$, longitude of ascending node $\Omega$, argument of pericenter $\omega$, and time of pericenter passage $T_p$), that can be either heliocentric or barycentric, defined at an arbitrary epoch $t$, for multiple object populations. These can range from near-circular ($e\approx 0$) orbits in the case of, for example, MBAs or TNOs (trans-Neptunian objects), parabolic ($e=1$) for long-period comets, or hyperbolic ($e>1$) for interstellar objects (ISOs). To be able to handle all of these scenarios efficiently, we implement the universal variables method of \cite{stumpff1947}, as described in \citet{Danby.1992}. The primary advantage of this methodology is that a single procedure can be used to handle all transformations, independent of the (physically allowed) value of $e$. 

We test our implementation in a few different ways. First, we begin with a toy model, in a unitless coordinate system in which the reduced mass $\mu = 1$. Any orbit in this system with $i = \Omega = \omega = 0\degr$ and is at perihelion ($T_p = t$ or $\mathcal{M} = 0 \degr$) will have $x = q$ and $y = z = 0$. Such an orbit also has $v_x = v_z = 0$, and $v_y > 0$ by construction. We verify that our transformations match these criteria by choosing a unitless value of $q=10$ and numerically computing these transformations for $0 \leq e \leq 10$, including $e=1$ and small perturbations around this neighborhood, with differences in the predicted positions and the expected values smaller than $10^{-15}$. We also verify that by changing $i$, $\Omega$, and $\omega$, the orbit is rotated accordingly.

We then proceed to test our transformations by comparing our state vectors to those provided by JPL Horizons for three different objects: main belt asteroid (3666) Holman, ISO 1I/\oumuamua, and the eccentric TNO 2014 WB$_{556}$. We obtain differences on the order of $10^{-10} \, \mathrm{au}$ in the derived positions for \oumuamua, and $10^{-13} \, \mathrm{au}$ for the other two objects, and, for all three objects, velocity differences are smaller than $10^{-13} \, \mathrm{au}/\mathrm{day}$.

\subsection{\texttt{ASSIST} Integrator}\label{sec:assist_integrator}

For accurate positions of the target objects, we use \texttt{ASSIST}~\citep{Holman.2023}, a package for carrying out ephemeris-quality integrations of test particles. 
By design, the \texttt{ASSIST} integrator almost identically matches the results of the JPL small bodies integrator, the current standard of reference for small body ephemerides.  \texttt{ASSIST} is an extension of \texttt{REBOUND}~\citep{rein2012} and makes use of its IAS15 integrator~\citep{rein2015} to integrate test particle trajectories in the field of the Sun, Moon, planets, and 16 massive asteroids, with the positions of the masses coming from the JPL DE440/441 ephemeris and its associated asteroid perturber file~\citep{Park.2021}. ASSIST incorporates the most significant gravitational harmonics and general relativistic corrections. It can also account for position- and velocity-dependent non-gravitational effects. 

Although \texttt{ASSIST} integrations are fast, we exploit a feature of the IAS15 integrator to further optimize ephemeris calculations.   An individual IAS15 time step involves iteratively solving for the coefficients of a 15-th order polynomial that accurately approximates the trajectory over the interval from the start of the time step to the end~\citep{rein2015}.   After converging, the  polynomial is used to evaluate the location of the object at the end of the step.  However, the same polynomial coefficients can be used to evaluate the location of the object at any time within the time step interval.  The high order of the polynomial enables each time step to be quite large (several days for MBAs, longer for TNOs and shorter for NEOs).  Evaluating the polynomial is computationally negligible compared to the integration step itself.  

\texttt{REBOUND} retains the coefficients of the previously completed time step.  \texttt{ASSIST} includes a routine called {\textit integrate\_or\_interpolate} that accepts a time as an argument.  If that time lies within the interval of the previous time step, the routine interpolates using the existing coefficients.  Otherwise, it takes the required time step.  
The nominal LSST observing strategy involves taking three pairs of exposures on distinct nights spanning 10-15 days.  The typical interval between successive exposures of a region of sky is often smaller than an IAS15 time step, so most ephemeris calculations can be done with polynomial evaluations.  
An evaluation of the interpolation polynomials for a single test particle is $\sim30,000$ times faster than a single full integration step.   The degree to which this optimization speeds up any particular \sorcha simulation depends upon the details of the survey.  If there are many visits of the same region of sky within a relatively short span of time, the speed up can be significant.  However, for surveys with exposures widely separated both spatially and temporally, the speed up from interpolating will be limited.  For typical solar system surveys, a speed up of a factor of 2-4 can be expected from using using this optimization.

Numerical integration with \texttt{ASSIST} has been extensively tested~\citep{Holman.2023}.  Nevertheless, we examine a few cases to demonstrate that we are using \texttt{ASSIST} correctly within \sorcha.  We do this by comparing the state vectors from \texttt{ASSIST} with those from JPL Horizons after a month long integration (starting at epoch 2017 October 1st, midnight TDB), for a few different objects. Our targets include objects in near circular orbits at a variety of heliocentric distances (including an NEO), objects with high eccentricities (a TNO, 2014 WB$_{556}$ with barycentric $e \approx 0.85$ and a long-period comet (LPC), C/2014 UN$_{271}$ with barycentric $e \approx 0.9975$), and an ISO (eccentricity $e > 1$).  We present the fractional differences between the reference values for JPL state vectors and astrometry (from Rubin Observatory) and those derived by \sorcha  in Table \ref{table:reference}. We also cross-validated the ephemeris of (3666) Holman against the independent software package \texttt{spacerocks}\footnote{\texttt{spacerocks} is available online at \url{http://github.com/kjnapier/spacerocks}}, and found excellent agreement at the level of tens of microarcseconds. 

\begin{deluxetable}{cc|cccc}[ht!]
	\tabletypesize{\footnotesize}
	\tablecaption{State vector and astrometry differences after month-long integrations\label{table:reference}}
	\tablehead{\colhead{Reference object} & \colhead{Orbit type} & \colhead{${|| \mathbf{r}_\mathrm{Sorcha} - \mathbf{r}_\mathrm{JPL}||}/{||\mathbf{r}_\mathrm{JPL}||}$} & \colhead{${|| \dot{\mathbf{r}}_\mathrm{Sorcha} - \dot{\mathbf{r}}_\mathrm{JPL}||}/{||\dot{\mathbf{r}}_\mathrm{JPL}||}$} & \colhead{$\Delta \alpha \cos \delta$} \tablenotemark{a}& \colhead{$\Delta \delta$}}
	\startdata
    2010 TU$_{149}$ & MBA & $5.3 \times 10^{-8}$ & $1.2 \times 10^{-8}$ & $-26 \, \mu \mathrm{as}$ & $-3.2\, \mu \mathrm{as}$\\ 
    (3666) Holman & MBA & $5.9 \times 10^{-8}$ & $2.0 \times 10^{-8}$ & $-23 \, \mu \mathrm{as}$ & $5.8 \, \mu \mathrm{as}$\\
    2011 OB$_{60}$ & TNO & $4.3 \times 10^{-8}$ & $6.2 \times 10^{-7}$ & $0.7 \, \mu \mathrm{as}$ & $0.9 \, \mu \mathrm{as}$ \\ 
    2014 WB$_{556}$ & TNO & $4.3 \times 10^{-8}$ & $7.8 \times 10^{-7}$ & $2.2 \, \mu \mathrm{as}$ & $1.6 \, \mu \mathrm{as}$\\
    1I/\oumuamua & ISO & $9.6 \times 10^{-5}$ & $3.1 \times 10^{-4}$ & $-24''$ & $-0.89''$\tablenotemark{b}\\ 
    (433) Eros & NEO & $6.8 \times 10^{-8}$ & $1.1 \times 10^{-8}$ & $-23 \, \mu \mathrm{as} $ & $5.8 \, \mu \mathrm{as}$\\ 
    C/2014 UN$_{271}$ (Bernardinelli-Bernstein) & LPC & $4.3 \times 10^{-8}$ & $3.0 \times 10^{-7}$ & $0.7 \, \mu \mathrm{as} $ & $1.3 \, \mu \mathrm{as} $
	\enddata
    \tablenotetext{a}{The RA/Dec positions were calculated with respect to Rubin Observatory.}
    \tablenotetext{b}{1I/\oumuamua's ephemerides from JPL includes non-gravitational acceleration terms that are not captured in our simulations.}
\end{deluxetable}

\subsection{Light Time Iteration}

The time it takes for the light from a solar system object to reach the observer can be significant.  During that time, the object will have moved appreciably.  Thus, the observer is seeing where the object was when the observed light was reflected by or emitted from the object.  In order to calculate the light time, one must know the position of the object, but to calculate the position of the object one must know the light time.  Thus, equation~\ref{Eq:light-time} is implicit.  

Fortunately, iteration is a simple and reliable approach to solving for the light time.  
One starts with an initial guess for $\Delta t$, such as $\Delta t=0$, finds $\rho =|\mathbf{r}(t-\Delta t) - \mathbf{r_{obs}}(t)|$, and updates $\Delta t = \rho/c$.  This procedure typically converges within a few iterations. 
After the light time iteration, the ephemeris position at time $t$ is known.

Rather than testing for convergence, we set the default number of iterations to three. 
We verified that changing the number of iterations from three to four iterations does not appreciably change the results, implying convergence. As a final check, we verified that $\Delta t = |\mathbf{r}(t-\Delta t) - \mathbf{r_{obs}}(t)|/c$.

\subsection{Validation of Simple Ephemeris Generation} \label{sec:validation_simple}

We have tested each of the steps separately, and we have verified the overall calculation using the same objects as in Section \ref{sec:assist_integrator}. If all previous steps have been done correctly, our predicted on-sky positions from the Rubin Observatory's location should closely match those produced by JPL. We start with our state vectors on a certain date and produce ephemerides for observations a month later. The results, presented in Table \ref{table:reference}, show remarkable agreement (at most tens of micro-arcseconds for objects with purely gravitational motion) between our integrations and JPL's across a variety of heliocentric distances, thus validating our ephemerides generation procedure. 

\section{Large-scale Ephemeris Generation} \label{sec:large-scale}

Although simple ephemeris generation is efficient enough for many purposes, the brute-force approach becomes a computational bottleneck for survey simulations at LSST scale.  As already noted (Section \ref{sec:intro}), we are considering millions of objects and millions of exposures, with the goal of determining which objects appear within the FOV of the survey camera for each exposure.  Calculating the ephemeris positions of all of the objects at the times of all of the visits becomes prohibitively expensive, even with significant computing resources.  Fortunately, two simple insights, when combined, lead to a much more efficient method: 
\begin{enumerate}
    \item The ephemeris positions of most objects are far outside the field of view of any particular exposure.  Even an approximate position is sufficient to determine that an object cannot plausibly fall within the field of view.    The $9.6 \, \deg^2$ field of view of LSSTCam, although enormous by current standards, only encompasses $\sim 1/4000$ of the sky.  Even near the ecliptic, only a tiny fraction of the total sample of objects in a survey simulation will appear in any given exposure.
    \item  The apparent rates of motion of most solar system objects, as observed from Earth, are slow.  Although NEOs can move several degrees across the sky per day, most MBAs traverse only a fraction of a degree in a day.  The apparent motion of more distant objects (e.g., TNOs) is even slower.  Thus, an object observed on one night will be in roughly the same location the next.
\end{enumerate}  

\subsection{High-level Algorithm}

\begin{figure}
    \centering
    \includegraphics[width=0.7\columnwidth]{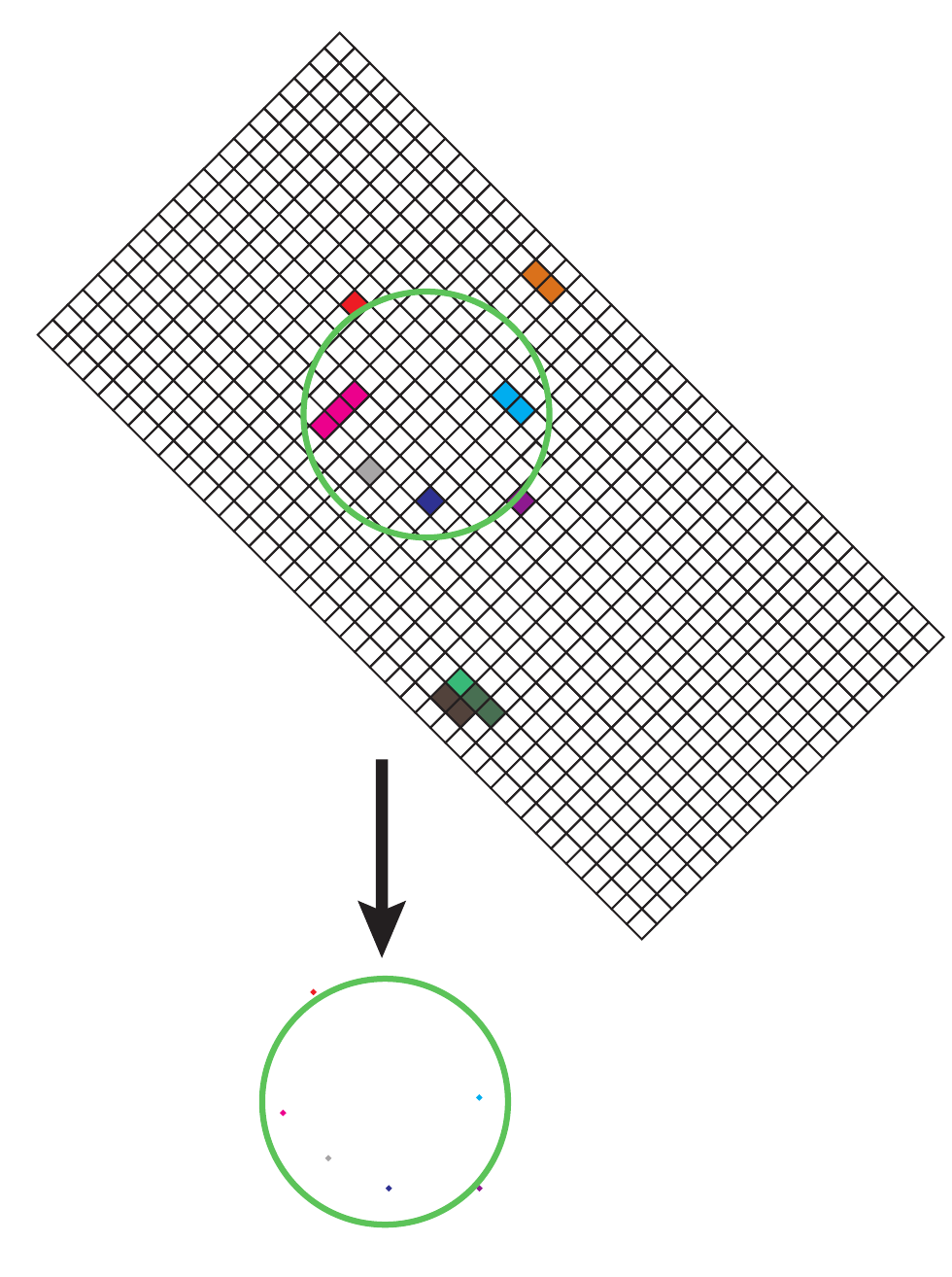}   
    \caption{Cartoon schematic of ephemeris generation within \sorcha.  Each color represents a different moving object on a different orbit. Slower moving objects will cover fewer HEALPix tiles. The green circle represents an area slightly bigger than LSSTCam's footprint. For the given observation time, any orbits with a HEALPix tile within the circle are integrated to calculate their exact positions at the time of the observation. Those orbits that land within the circle are then identified, and the resulting ephemerides associated with those objects and the observation are saved and passed on to later stages of the simulator that calculate the apparent magnitude and estimate the likelihood of discovery.}
    \label{fig:cartoon_ephem_gen}
\end{figure}

Merging these two ideas yields our basic approach (demonstrated in the Figure \ref{fig:cartoon_ephem_gen}).  We build and periodically update a data structure that allows us to efficiently determine which simulated objects are in which region of sky, on a given night. 
We call this data structure, which records the topocentric positions for all the objects, a \emph{picket}, because we will be computing a sequence of these, like pickets in a fence.  Although one could store these pickets, we will maintain one at a time, to limit the memory footprint of the simulation.  

We use the HEALPix\footnote{HEALPix: Hierarchical Equal Area Iso Latitude pixelation} tessellation~\citep{Gorski.2011}, which divides the sky into equal-area pixels and enumerates them.  (We orient the HEALPix tessellation with the ICRF and uses the $n_{side}=128$ scale\footnote{The HEALPix $n_{side}$ scale value is configurable within \sorcha, but we have found a default of $n_{side}=128$, which corresponds to a scale of $0.45$ degrees, worked well for all known solar system populations. }.) 
For a given night, our data structure maintains a mapping between the HEALPix index and the set of simulated objects that fell within the corresponding region of the sky at any point during the night.

Given the RA/Dec of a given exposure, we find all the HEALPix tiles overlapped by a circular region with an angular radius that encompasses the camera FOV, including a buffer region to accommodate some uncertainty (discussed below).  We then collect the IDs of the simulated objects in those tiles and calculate precise ephemeris positions for the exposure time for these candidate objects.  Any objects that actually fall within the angular radius of the field of view are retained.
 
Other tessellations of the sky would also work.  The only requirements are that it is easy to (1) determine the tile index from a sky position and (2) determine which tiles are overlapped by the FOV of an exposure.  We selected HEALPix because the \emph{healpy}\footnote{\url{https://healpy.readthedocs.io/en/latest/}} \citep{Gorski.2011,zonca2019} provides efficient Python support for these operations. 

There are many ways to implement and exploit the aforementioned concepts.  We chose to employ the following procedure, but we also note that further refinements are possible.
Consider a toy example in which we have a single exposure, and we wish to determine which of the simulated objects fall within its bounds.  The exposure has a time, a topocentric unit vector pointing toward the center of the FOV, and an angular radius that circumscribes the FOV.  The FOV of the exposure overlaps a set of HEALPix tiles, which we can determine in advance.

Within \sorcha, for each object in our simulated population:
\begin{enumerate}
    \item Calculate the ephemeris positions on three sequential nights (see subsection~\ref{sec:ephemeris}), ensuring that the epoch of the exposure is within that time span. 
    \item Finely interpolate those topocentric positions across that two-day interval.
    \item Determine which HEALPix tiles the object will traverse in that time interval. 
    \item Associate the object's ID with the indices of the HEALPix tiles it crosses (we use a hash table/dictionary for this purpose).
\end{enumerate}

 Then obtain all objects that could possibly appear within the exposure by collecting the set of object IDs in the set of HEALPix tiles that the exposure overlaps.  For each of those candidate objects, determine its precise ephemeris position at the time of the exposure.  Only those that actually fall within the FOV are retained.
 
For a simulation of an LSST survey with millions of exposures, part of the advantage of using this approach is that steps 1-4 do not necessarily need to be repeated for each exposure.  Consider a time-ordered sequence of exposures spanning a number of days.  After calculating steps 1-4 for the first exposure, the results can be reused for all the exposures whose epochs land within the two-day interval spanned by the three points in the picket.  If an exposure falls outside that interval, the data must be updated.


\subsection{Orbits and Ephemerides}~\label{sec:ephemeris}

The toy example in the previous subsection did not elaborate on how the propagation of orbits is done, but it is computationally intensive and extremely difficult to execute correctly.
In order to allow our simulations to scale well to the volume of LSST data, we need to be able to handle millions of orbits.  To achieve the fidelity required by a survey that spans a decade and has superb astrometric precision, we need to use an extremely accurate integration scheme that supports high-quality ephemerides.   The ephemeris generation tool needs to be able to achieve this with a reasonable number of computing cores and memory in a reasonable amount of time.  

To enable this performance and precision, we have taken a counterintuitive approach in which we input the initial orbits one by one and instantiate an individual \texttt{REBOUND}/\texttt{ASSIST} simulation object for each orbit.
Since we treat the small bodies as massless test particles (i.e. they do not interact with each other), it would seem more efficient to incorporate a large number of solar system bodies into each numerical integration.  For fixed time step integrators, particularly when the orbits of the planets and other perturbing bodies must also be numerically integrated, one can amortize the computational overhead by applying the same calculation to a large number of test particles.  
However, \texttt{ASSIST} does not numerically integrate the positions of the Sun, planets, and other perturbing bodies.  Instead, it efficiently evaluates them through memory-mapped tables of Chebyshev polynomial coefficients.  Furthermore, integrating groups of test particles is only efficient if the step size needed is roughly the same for all of the test particles.  If a few require particularly small time steps in order for their results to be accurate, integrations of the other objects may be unnecessarily expensive.  
This is also true for IAS15, despite it being a variable time step method.  Each simulation has a single, variable time step that is applied to all of the equations in the simulation.  The length of the time step is dominated by the most rapidly changing equation~\citep{rein2015}.  Solar system objects have a wide range of orbital periods that correspondingly require a wide range of step sizes.  By separating each orbit into its own REBOUND simulation, the integrator will naturally use the time step that is best suited to that object.  Furthermore, this means that each object has its own set of polynomial coefficients that can be used to evaluate its position within a time step.  
In addition, keeping a separate \texttt{REBOUND}/\texttt{ASSIST} object for each orbit allows for easy bookkeeping: as we read in the orbits and instantiate the REBOUND/ASSIST objects, we build a python dictionary that has the object IDs as keys and the REBOUND/ASSIST objects as values.

\subsection{Processing Exposures}

Next, we process the survey exposures, one by one, roughly following the approach described in the toy example above.  Although the exposures do not need to be processed in chronological order, it is far more efficient to do so. 
We process  the exposures as follows:
\begin{enumerate}
\item Read the exposure mid-point time and RA/Dec of the survey pointing center.  Calculate the observatory position for that time, if it is not already available.

\item Calculate which HEALPix tiles the exposure overlaps.  

\item If the time of the exposure is outside of the time interval of the current picket, update the picket.

\item Collect the IDs of the objects in the HEALPix tiles that the exposure overlaps.

\item Integrate just those objects to the time of the exposure, including light time.  
\item After the set of objects that \emph{might} land in the exposure have been integrated, check which actually fall within the angular radius of the FOV from the center of the exposure.  For those objects, calculate and record the rates of motion and any other information that is needed in the post-processing stages of \sorcha.

\end{enumerate}
This process is essentially implemented as a generator, hiding the details from the user.  For a given exposure, the process returns the information for the objects that fall within that exposure.

As noted earlier (Section \ref{sec:assist_integrator}), the  \textit{integrate\_or\_interpolate} routine makes this step much more efficient.  The time steps can be large, often several days.  Thus, most of the time no actual numerical integration needs to be done.  In addition, because each of the test particles has its own REBOUND/ASSIST object, the adjustment of the integration time steps is handled simply and automatically.

There will be hundreds to thousands of exposures per night of LSST observations.  If the exposures are processed in time order,  much of the time the picket does not need to be updated.  Sequences of exposures taken within the same night or adjacent nights can use the same picket.  Groups of exposures separated by more than a day will typically need different picket information.  The interval between pickets is discussed in more detail below.  \deleted{Likewise, few test particles will need to be integrated to process a given exposure.  The times of many of the exposures will fall within the limits of an integration time step, allowing polynomial evaluation of the positions.}

\subsection{Design Considerations}
The design of the picket data structure and its associated algorithms was motivated by fast-moving objects.  NEOs, for example, can traverse several HEALPix tiles between one night and the next, depending upon the topocentric distance to the object, the interval between pickets, and the scale of the HEALPix tiles.  We need to allow for this possibility to keep track of such objects.
Fortunately, there is a fairly simple solution to this problem.  When updating the picket, we record all the HEALPix tiles each object will cross in the two-night interval covered by the picket.  Then we add the object IDs to the hash map/dictionary for each of those HEALPix tiles.  We use moving-window third-order Lagrange polynomials to interpolate the topocentric unit vectors to the objects, across \deleted{the} adjacent three adjacent nights.  We evaluate the position (and HEALPix indices) at 101 interpolation points (this is adjustable via \sorcha's configuration file).  This adds negligible overhead to the overall calculation.

The interval between pickets is an adjustable parameter.  We set this to be one day to match the natural cadence of ground-based observations, and the amount of motion for the vast majority of objects in a typical simulation is modest on that time scale.  
Likewise, the HEALPix scale is a tunable parameter.  A smaller scale means that a set of pixels can more closely approximate the shape of the LSSTCam FOV.  However, a finer scale results in larger data structures and a larger memory footprint.  Although our testing demonstrates that these are reasonable values, we leave a detailed study involving the picket interval, the HEALPix scale, and the number of interpolation points to future work.

\section{Validation} \label{sec:validation}

We compare the results of \sorcha's optimized ephemeris generation with those of brute force application of the method described in Section~\ref{sec:simple}.  In particular, we aim to ensure that we are not missing objects. As part of our validation of \sorcha, described in detail in Merritt et al. (2025, submitted), we have simulated the observability of three objects (the NEO 433~Eros, the TNO 2011~OB$_{60}$, and the MBA 2010~TU$_{149}$)
over a month of LSST observations with a realistic cadence ($\approx 20,000$ visits), both through \sorcha and by directly producing ephemerides by querying JPL. These tests validate not only our ephemerides, but also the HEALPix tiling, the time picket algorithm, and the FOV selection criteria. The JPL-produced ephemerides have a cadence of a minute, and so we linearly interpolated these dense sets of positions to the corresponding exposure midpoint times from the simulated LSST visits, and used a simple independent implementation of the LSST FOV filtering for each visit. We obtain the same astrometric precision shown in Table \ref{table:reference} for these two objects for all detections over the chosen time span, and furthermore, these tests also indicate that all observations are properly identified by our algorithm.

\added{
We carried out a number of timing tests. First, we ran 10 MBA from the \sorcha demonstration files.  Using \sorcha with all the optimizations described in this paper, it takes 12 seconds, on a Macbook Pro with an M4 processor, for $\sim216,000$ visits over the span of a year of LSST observations.   Using the same input orbits and the same sequence of visits, but requiring the positions of all the objects to be calculated for each exposure time (i.e. a brute force calculation), takes 817 seconds on the same hardware.  

Second, we ran 100 Main Belt asteroids for $\sim216,000$ visits, on the same hardware.  This took 15 seconds with all optimizations, but 1,600 seconds as a brute force calculation.  The timing is roughly proportional to the number of objects simulated, with some overhead.

Third, again on a Macbook Pro, with an M4 processor, we ran $15,000$ objects for the same $\sim216,000$ visits.  We found that using {\textit integrate} took 745 seconds, while using {\textit integrate\_or\_interpolate} took 278 seconds.  This supports our earlier statement that in typical surveys, the use of {\textit integrate\_or\_interpolate} can yield a speed of 2-4.
}

\section{Caveats} \label{sec:caveats}
There are a few caveats regarding ephemeris generation that the \sorcha user should be aware of:
\begin{itemize}
    \item  Although we have tested the behavior of routines for very fast-moving objects, it is difficult to prove correctness in all cases.
    \item   If one of the objects listed in Table~\ref{tab:perturber} is included as a test particle in a \sorcha simulation, the results can be wildly inaccurate.  Without modifying the list of perturbers used by ASSIST, the test particle will experience the gravitational influence of a massive version of itself, at very close range.
    \item Currently, ASSIST includes only one formulation of non-gravitational acceleration. Likewise, as noted in Merritt et al. (submitted), ASSIST does not include routines for the acceleration due to cometary outgassing, the fragmentation of objects, monitoring for planetary impacts, and other effects that might be of interest to the community.   These and other features are feasible to account for, but these would need to be implemented in the software.  A primary motivation for making ASSIST and \sorcha open source projects is to enable other members of the community to develop and contribute such improvements.
    \item  For this version of \sorcha, we ignore orbital uncertainties and their resulting positional uncertainties.  For a survey simulation, we assume we know the orbits and physical characteristics of the sample of solar system objects being tested.  However, in future work, we can examine how positional uncertainties affect discovery, particularly in the early stages of the survey when there is limited data.
\end{itemize}

\begin{deluxetable}{l||l}
\tablecaption{The 16 MBA and dwarf planet perturbers included within the ASSIST integrations \label{tab:perturber}}
\tablehead{
\colhead{Asteroid MPC Designation } &  \colhead{Asteroid MPC Designation }
}
\tablecolumns{2}
\tablewidth{0pt}
\startdata
(107) Camilla &  (704) Interamnia \\
(1) Ceres  & (7)  Iris \\ 
(65) Cybele  & (3) Juno \\ 
 (511) Davida   & (2) Pallas \\
 (15) Eunomia  &  (16) Psyche \\
(31) Euphrosyne  &  (87) Sylvia \\
  (52) Europa  &  (4) Vesta   \\ 
 (10) Hygiea & (88) Thisbe
\enddata
\tablecomments{As these objects and their masses are included with the ASSIST integrations, they will gravitational perturb any orbits on similar or exact orbits. Thus, poor predictions will result if \sorcha ephemeris generator attempts to run on the perturber orbits.}
\end{deluxetable}

\section{Summary} \label{sec:summary}

The dawn of the petabyte era in astronomical surveys brings great challenges and rewards. The significant increase in photometric and astrometric observations requires more efficient algorithms to  accurately predict which orbits will land within past survey images. This is both to identify where known the millions of known minor planets recorded in the MPC have serendipitously been imaged by these large surveys, but also to enable simulations that predict the locations of millions to billions of synthetic objects in survey observations in order to compare population models to that discovered by these surveys. We have described the fast ephemeris generation algorithm designed for the \sorcha survey simulator (Merritt et al. submitted) and how we have overcome many of the challenges of scaling up for the LSST.  We note that the ephemeris generator within  \sorcha has been designed and coded with ASSIST and REBOUND in mind, but the algorithm itself is agnostic to the choice of n-body integrator.  Likewise,  \sorcha's ephemeris generator can be used separately from the rest of the package.  For example, the ephemeris generation components have been integrated into the known object association software within the Rubin Observatory's Solar System Processing pipelines.\footnote{https://dmtn-087.lsst.io/}
Our algorithm may also be of particular interest to large astronomical databases such as the archives at MAST\footnote{\url{https://archive.stsci.edu}} (Barbara A. Mikulski Archive for Space Telescopes) and the Astro  Data Lab\footnote{\url{https://datalab.noirlab.edu/}} at the National Optical-Infrared Astronomy Research Laboratory (NOIRLab).

Future work could include (1) propagating positional uncertainties as part of ephemeris generation, (2) efficiently storing the data at a series of pickets to support the efficient identification of known small bodies for a range of times, (3) a detailed study involving the picket interval and HEALPix scale, as noted in section~\ref{sec:large-scale}, and (4) adding the capacity to keep a short list of objects that are always evaluated via brute force methods, to guarantee that key objects are not missed.

\vspace{5mm}

\software{sorcha, ASSIST \citep{Holman.2023,rein2023}, Astropy \citep{astropy2013,astropy2018,astropy2022}, Healpy \citep{Gorski.2011,zonca2019}, Numba \citep{lam2015}, Numpy \citep{harris2020}, pandas \citep{mckinney2010, pandas2020}, REBOUND \citep{rein2012, rein2015}, SciPy \citep{virtanen2020}, Spiceypy \citep{annex2020}, Black (\url{https://black.readthedocs.io/en/stable/faq.html}), Jupyter Notebooks \citep{kluyver2016}, spacerocks (\url{https://github.com/kjnapier/spacerocks}}.

\section*{Author Contributions}

M.J.H. led the development and refinement of Sorcha's ephemeris generator algorithm.  He also helped verify the output from the ephemeris generator components and improved the algorithm to better handle fast-moving solar system objects. He also wrote the majority of the text within the manuscript. 

P.H.B. contributed to the development of several of the \sorcha routines for ephemerides generation and coordinate conversion. He also helped verify the output from the ephemeris generator components and improved the algorithm to better handle fast-moving solar system objects. He also contributed significant portions of text to the manuscript. 

M.E.S. served as principal investigator and project manager of the \sorcha team. She also served as PI of the \sorcha LINCC Frameworks Incubator Proposal. She contributed to the discussions and decisions about the overall design and implementation of \sorcha and the documentation: this included leading the team in-person meetings and calls. She created \ref{fig:cartoon_ephem_gen}, contributed some text to the paper, and also provided feedback on the manuscript. 

M.J. contributed to the original brainstorming sessions about the \sorcha ephemeris generator design and implementation. He wrote the original Jupyter  notebook that was used as a template by M.J.H to incorporate ASSIST+REBOUND. M.J. also searched through a petabtye of files on disk to provide the notebook to M.J.H. 

D.O. led the incubator software engineering team during the LINCC Frameworks Incubator and contributed to integrating the first draft of the ASSIST+REBOUND ephemeris generator within \sorcha. 

M.W. contributed to the \sorcha code base, including efficiency improvements in array handling and integrating the first draft of the ASSIST+REBOUND ephemeris generation.

K.J.N. cross-validated the \sorcha ephemeris against the ephemeris produced by the independent software \texttt{spacerocks} (\url{https://github.com/kjnapier/spacerocks}). He also helped to review parts of the software, and identified minor algorithmic improvements.

S.R.M. served as the lead developer of \sorcha from June 2021 onward, contributing a substantial portion of the current code base and the majority of unit tests. They also led new feature implementations, oversaw ongoing code maintenance and repository management, resolved software issues, conducted extensive testing on both HPC clusters and local systems.

G.F. was the initial developer of \sorcha and contributed significantly to early drafts of the manuscript.  G.F. participated in the discussions and decisions about the overall design and implementation of \sorcha. He also provided feedback on the overall paper draft. 

S.C. contributed to \sorcha and assisted in stress testing \sorcha's ephemeris generator and identifying rare cases during these beta tests that required changes that improved the ephemeris generation within \sorcha. 

J.A.K. provided testing, user feedback, and bug-fixing throughout \sorcha's development. He also provided feedback on the paper manuscript.

S.E. contributed to the initial discussions about  \sorcha's ephemeris generator as well as providing feedback on the LINCC Frameworks Incubator proposal. He was involved in the discussions about the overall design and functionality of \sorcha.  He also provided feedback on the paper manuscript. 

J.K. contributed to the \sorcha code base, including improvements to input file reading of the user generated ephemeris files.

K.K. contributed to beta testing \sorcha for the edge case of simulated Earth impactors and very fast-moving potentially hazardous asteroids (PHAs). 

J. M. provided testing, user feedback, and bug-fixing throughout \sorcha's development. He also provided feedback on the paper manuscript.

S.P.N. developed the \texttt{objectsInField} (\texttt{OIF}) ephemeris generator/simulator that enabled the development of \sorcha and its epehmeris generator. The output from OIF was the input into early iterations of \sorcha as the source of ephemeris calculations. OIF's design influenced the overall architecture of the \sorcha ephemeris generator.  

C.O.C. contributed to discussions about the development and enhancement of \sorcha, including the ephemeris generator, during the LINCC Frameworks Incubator. C.O.C. also provided feedback on the paper manuscript.

\section*{Acknowledgments}

 This work was supported by an LSST Discovery Alliance LINCC Frameworks Incubator grant [2023-SFF-LFI-01-Schwamb]. Support was provided by Schmidt Sciences.  M.J.H. gratefully
acknowledges support from the NSF (grant No. AST2206194), the NASA YORPD Program (grant No.
80NSSC22K0239), and the Smithsonian Scholarly Studies Program (2022, 2023).   S.R.M. and M.E.S. acknowledge support in part from UK Science and Technology Facilities Council (STFC) grants ST/V000691/1 and ST/X001253/1. G.F. acknowledges support in part from UK Science and Technology Facilities Council (STFC) Grant ST/P000304/1. This project has received funding from the European Union’s Horizon 2020 research and innovation programme under the Marie Sk\l{}odowska-Curie grant agreement No. 101032479. C.O.C, M.J. and P.H.B. acknowledge the support from the University of Washington College of Arts and Sciences, Department of Astronomy, and the DiRAC (Data-intensive Research in Astrophysics and Cosmology) Institute. The DiRAC Institute is supported through generous gifts from the Charles and Lisa Simonyi Fund for Arts and Sciences and the Washington Research Foundation. M.J. wishes to acknowledge the support of the Washington Research Foundation Data Science Term Chair fund, and the University of Washington Provost's Initiative in Data-Intensive Discovery. J. M. acknowledges support from the Department for the Economy (DfE) Northern Ireland postgraduate studentship scheme and travel support from the STFC for UK participation in LSST through grant ST/S006206/1. J.A.K.  and J. M.  thank the LSST-DA Data Science Fellowship Program, which is funded by LSST-DA, the Brinson Foundation, and the Moore Foundation; their participation in the program has benefited this work. S.E. and S.C. acknowledge support from the National Science Foundation through the following awards: Collaborative Research: SWIFT-SAT: Minimizing Science Impact on LSST and Observatories Worldwide through Accurate Predictions of Satellite Position and Optical Brightness NSF Award Number: 2332736 and Collaborative Research: Rubin Rocks: Enabling near-Earth asteroid science with LSST NSF Award Number: 2307570. Any opinions, findings, and conclusions or recommendations expressed in this material are those of the authors and do not necessarily reflect the views of the National Science Foundation.
 
 This work was also supported via the Preparing for Astrophysics with LSST Program, funded by the Heising Simons Foundation through grant 2021-2975, and administered by Las Cumbres Observatory. This work was supported in part by the LSST Discovery Alliance Enabling Science grants program, the B612 Foundation, the University of Washington's DiRAC Institute, the Planetary Society, Karman+, and Adler Planetarium through generous support of the LSST Solar System Readiness Sprints. 

 This research has made use of NASA’s Astrophysics Data System Bibliographic Services. This research has made use of data and/or services provided by the International Astronomical Union's Minor Planet Center. The SPICE Resource files used in this work are described in \citet{acton1996, acton2018}. Simulations in this paper made use of the REBOUND N-body code \citep{rein2012}. The simulations were integrated using IAS15, a 15th order Gauss-Radau integrator \citep{rein2015}. Some of the results in this paper have been derived using the \texttt{healpy} and HEALPix packages. This work made use of Astropy:\footnote{http://www.astropy.org} a community-developed core \python package and an ecosystem of tools and resources for astronomy \citep{astropy2013,astropy2018,astropy2022}. We are additionally grateful to the members of the Rubin Observatory LSST Solar System Science Collaboration for useful feedback at the LSST Solar System Readiness Sprints. We also thank the contributors to Stack Overflow for their examples and advice on common \python challenges that provided guidance on solving some of the programming challenges we have encountered. 

 This material or work is supported in part by the National Science Foundation through Cooperative Agreement AST-1258333 and Cooperative Support Agreement AST1836783 managed by the Association of Universities for Research in Astronomy (AURA), and the Department of Energy under Contract No. DE-AC02-76SF00515 with the SLAC National Accelerator Laboratory managed by Stanford University.  

We are grateful for use of the computing resources from the Northern Ireland High Performance Computing (NI-HPC) service funded by EPSRC (EP/T022175). We gratefully acknowledge the support of the Center for Advanced Computing and Modelling, University of Rijeka (Croatia), for providing supercomputing resources at HPC (High Performance Computing) Bura.

Data Access:  The algorithm presented here is available in the \sorcha open-source \python package at \url{https://github.com/dirac-institute/sorcha}.

\bibliographystyle{aasjournal}

\end{document}